# TO PARALLELIZE OR NOT TO PARALLELIZE, CONTROL AND DATA FLOW ISSUE

A. I. El-Nashar

Computer Science Department, Faculty of Science, Minia University,
Minia - Egypt
nashar_al@yahoo.com

**Abstract:** New trends towards multiple core processors imply using standard programming models to develop efficient, reliable and portable programs for distributed memory multiprocessors and workstation PC clusters. Message passing using MPI is widely used to write efficient, reliable and portable applications. Control and data flow analysis concepts, techniques and tools are needed to understand and analyze MPI programs. If our point of interest is the program control and data flow analysis, to decide to parallelize or not to parallelize our applications, there is a question to be answered, " Can the existing concepts, techniques and tools used to analyze sequential programs also be used to analyze parallel ones written in MPI?". In this paper we'll try to answer this question.

## 1. Introduction

A concurrent program contains two or more threads that execute concurrently and work together to perform some task [17]. Using multiple threads can increase the efficiency. The increase in the use of parallel computing is being accelerated by the development of various parallel hardware architectures and also by constructing new programming models to achieve the best use of these architectures. Message Passing Interface or MPI is one of these models which is widely used for programming distributed memory systems. While concurrent programs offer some advantages, they also exhibit nondeterministic behavior, making them difficult to test. One significant challenge in bringing the power of parallel machines to application programmers is providing them with a suite of software tools similar to the tools that sequential programmers currently utilize as a practical way of obtaining increased confidence in software and also to guarantee that the program is correct. In particular, automatic or semi automatic testing tools for parallel programs are lacking compared with that tools for sequential programs [1].

Control flow and data flow analysis of serial programs differ from those of MPI parallel programs as a result of complex interaction between concurrent processes and also due to the SPMD nature. The overhead of analyzing this type of programs may make application programmers to think a lot to decide to parallelize or not to parallelize and as a result they may prefer using the ordinary serial programming style to be not faced with the extra analysis effort. This paper presents an implemented technique to analyze MPI programs. The analysis report serves as a guide to the programmer to compromise between the advantage of parallelism and the cost of analysis effort.

The paper is organized as follows: Section 2 describes the MPI Programming Model. In section 3, MPI Program Analysis is presented. Section 4, concerns with MPI Program Representation. MPI Static Data Flow Analysis technique is described in section 5.

## 2. MPI Programming Model

MPI programs are coded in a special manner, in which each process executes the same program with unique data. All parallelism is explicit; the programmer is responsible for correctly identifying parallelism and implementing parallel algorithms using MPI constructs.

```
    1    1           include 'mpif.h'
    1    2           integer ierr, myid, status(MPI_STATUS_SIZE)
    1    3           integer count,x,sum,received,sender,process
    1    4           SUM=3
    1    5           call MPI_Init(ierr)
    1    6           call MPI_Comm_size(MPI_COMM_WORLD, count, ierr)
    1    7           call MPI_Comm_rank(MPI_COMM_WORLD, myid, ierr)
    2    8           IF( MYID.EQ.0) THEN
    3    9              call MPI_Recv(received,1,MPI_INTEGER, MPI_ANY_SOURCE,
             $                MPI_ANY_TAG, MPI_COMM_WORLD, status, ierr)
    4   10              call MPI_Recv(sender, 1, MPI_INTEGER,MPI_ANY_SOURCE,
             $                MPI_ANY_TAG, MPI_COMM_WORLD, status, ierr)
    5   11              SUM=SUM+ RECEIV
    5   12              WRITE(*,5) RECEIVD , SENDER, SUM
    5   12 5      format (' The value: ', I3 , 2x , 'is received from process:',
             $              I3 , 'The sum:' , I3)
    6   13         ENDIF
    7   14         IF( MYID.EQ.1 ) THEN
    8   15            X=5
    8   16            IF(X.LT.0) THEN
    9   17               X=X+1
   10   18             ELSE
   10   19               X=X-1
   11   20            END IF
   12   21            call MPI_Send(X,1,MPI_INTEGER,0,0,MPI_COMM_WORLD,ierr)
   13   22            PROCESS=MYID
   14   23            call MPI_Send(process,1,MPI_INTEGER,0,0,MPI_COMM_WORLD,ierr)
   15   24         ENDIF
   16   25         IF( MYID.EQ.2 ) THEN
   17   26            X=7
   17   27            X=X * 2
   18   28               call MPI_Send(X,1,MPI_INTEGER,0,0,MPI_COMM_WORLD,ierr)
   19   29            PROCESS=MYID
   20   30            call MPI_Send(process,1,MPI_INTEGER,0,0,MPI_COMM_WORLD,ierr)
   21   31         ENDIF
   22   32         call MPI_Finalize(ierr)
   23   33      END
```

Figure 1. A typical MPI program

MPI is available as an open sources implementations on a wide range of parallel platform [6,9,15,16]. MPICH2 [4,10] is a recent MPI open sources implementation, in which the MPI source program is compiled and then linked with the MPI libraries to obtain the executable. The user issues a directive to the operating system that places a copy of the executable program on each processor, the number of processes is provided within the user directive. Each processor begins execution of its copy of executable. Each process can execute different statements by branching within the program based on a unique rank "process identifier". This form of MIMD programming is frequently called Single-program multiple-data SPMD. Each process has its own local memory address space; there are no shared global variables among processes. All communications are performed through special calls to MPI message passing routines.

MPI uses objects [11] called communicators and groups to define which collection of processes may communicate with each other. A communicator must be specified as an argument for most MPI routines. The predefined communicator MPI_COMM_WORLD is used whenever a communicator is required, it includes all of MPI processes. Within a communicator, every process has its own unique, integer identifier "rank" assigned by the system when the process initializes. Ranks are contiguous and begin at zero, used by the programmer to specify the source and destination of messages, and also used conditionally by the application to control program execution.

An MPI program consists of four parts, a typical MPI program is shown in figure 1. The first one is the MPI include file which is required for all programs/routines which make MPI library calls (line1). The second part is responsible for initializing MPI environment, MPI environment management routines are used for an initializing and terminating the MPI environment, querying the environment and identity. MPI_Init [8], (line 5), initializes the MPI execution environment. This function must be called in every MPI program, must be called before any other MPI functions and must be called only once in an MPI program.

MPI_Comm_size, (line 6), determines the number of processes in the group associated with a communicator. Generally used within the communicator MPI_COMM_WORLD to determine the number of processes being used by the application.

MPI_Comm_rank (line 7), determines the rank of the calling process within the communicator. Initially, each process will be assigned a unique integer rank between 0 and number of processes.
The third part is the body of program instructions, calculations, and message passing calls. The last one is terminating MPI environment (line 32). MPI provides several routines used to manage the inter-process communications via send / receive operations ( lines 9,10,21,23,28,30 ), wait for a message's arrival or probe to find out if a message has arrived.

## 3. MPI Program Analysis

Running the executable of the source code listed in figure 1 several times using three process may yields one of two outputs, one of them indicates that the value 4 is sent from process 1 to process 0 and the sum value is 7, the other one indicates that the value 14 is sent from process 2 to process 0 and the sum value is 17.

The order of these outputs is unpredictable. This situation reflects the non-deterministic behavior of program execution.

Applying ordinary data flow analysis that does not consider the SPMD nature of MPI programs on the source program listed in figure 1 to identify the statements influenced by the definitions of "sum" in lines 4 and 11 and also definition of "x" in lines 15,17,19,26 and 27 is shown in table 1 .

Table 1

| Case | Affected statements | |
|---|---|---|
| definition of sum in line 4 | 11 | sum=sum + received |
| definition of sum in line 11 | 12 | write(*,5) received , sender, sum |
| definition of x in line 15 | 16 | IF(x.LT.0) THEN |
| | 17 | x = x +1 |
| | 19 | x = x - 1 |
| definition of x in line 17 | 21 | call MPI_Send (x,…) |
| definition of x in line 19 | 21 | call MPI_Send (x,…) |
| definition of x in line 26 | 27 | x = x * 2 |
| definition of x in line 27 | 28 | call MPI_Send (x,…) |

The execution behavior demonstrates that the affected statements shown in table1 are not the only affected ones but there are some other statements that should be encountered, it can be noticed that this analysis fails to detect the effect of the definition of "x" on the computation of "sum" in line 11, as shown in table 2.

Table 2

| Case | Statements should be encountered |
|---|---|
| definition of x in lines 17, 19 | 9.   call MPI_Recv(received,…… |
| definition of x in line 27 | 9.   call MPI_Recv(received,…… |

Variable definitions like " sum = 3 " in line 4, are shared in SPMD programs without communication channels so they are considered as global variables. On the other hand, variables defined within each process section , enclosed between " IF( myid.eq.process_id) " and " ENDIF ", like " sum=sum + received" in line 11 can't be shared outside this section otherwise appropriate communication channels are used. These variables are considered as local variables.

## 4. MPI Program Representation

As shown above, ordinary data flow analysis techniques fail to demonstrate a correct analysis for MPI, the SPMD nature needs to be modeled correctly in the program representation to be considered during static program analysis, this can be achieved by using a special data structure that can represent sequential flow, parallel flow and synchronization in explicitly MPI parallel programs.

### 4.1 MPI-CFG Construction Challenges

Data flow analysis techniques represent a program by its control flow graph, CFG, which consists of a set of nodes and edges. Each node represents a basic block which is a set of consecutive statements of the program, and each edge represents the control flow between these blocks. The goal of analysis techniques is to identify which definitions of program variables can affect which uses. To build such CFG's, the analyzed program is divide into a set of basic blocks and the set of edges connecting these blocks according to the flow of control are generated. The constructed CFG is then used by the static analyzer to identify the def-use associations among the blocks.

Extra effort has to be done in building CFG representing MPI programs (MPI-CFG) according to the following challenges:

1. Processes in MPI programs are declared by using the ordinary conditional IF statement depending on a unique identifier. This will make a confusion during dividing the program into basic blocks, and also during the process of generating edges which will badly affect the operations of static analyzer. So, the IF statements that used to declare processes must be treated in a special manner rather than that is used in treating the IF statements used within the body of each process as shown in figure 1, lines 14 and 16 .
2. MPI program is executed by more than one process, each process has its local memory address space; there is no shared global variables among these processes except that are defined before the initialization of the MPI execution environment. This requires to identify both local and global variables.
3. def-use association among process can be achieved only by calling inter-process communication message passing routines (send, receive,…). This implies constructing extra edges that represent these constructs.

### 4.2 Implementation of MPI-CFG Construction

Now we present our algorithm to build the MPI-CFG. This flow graph resembles the synchronized flow graph [3], program execution graph [14] , parallel flow graph [7], and parallel program flow graph PPFG [2]. The algorithm works as follows:

1. **MPI program statements identification**.

   In this phase, each program statement is assigned a unique number "type" to be identified from the other statements of the program. The algorithm must check for the following:
   - If the statement " Call MPI_Comm_rank( XX,YY,ZZ) " is encountered, it is assigned its type and the second parameter YY which indicates the variable name that will be used to identify the parallel processes is recorded as "special_id".
   - The assigned type of conditional IF statements depends on the recorded "special_id"; if the value of "special_id" appears in the condition, this means that the encountered IF statement is used to declare a process, otherwise, it is an ordinary conditional statement.
   
   The output of this phase is a numbered statements of the input program associated with their types and the recorded "special_id".

2. **Building Basic Blocks**

   This phase uses the output of the previous phase to build the program basic blocks as in the case of ordinary CFG. We construct two extra special types of basic blocks called "message block" and "finalize block". A message block is either "receive block" or "send block". A basic block that has at most one communication statement at start of the block is said to be "receive block". This block is constructed if the statement call MPI_Recv( ) is encountered. The "send block" has at most one communication statement at its end. This block is constructed if the statement call MPI_Send( ) is encountered. The "finalize block" is constructed if call MPI_Finalize( ) statement is encountered.

During building basic blocks the program variables, their block numbers and their status (def, c-use, or p-use) are also recorded. Extra effort has to be done in case of call MPI_Recv( ) and call MPI_Send( ) to record and classify the parameters of these two calls. At the termination of this phase another version of the input MPI program is generated. This version contains the statement and block number for each program statement as shown in figure 1. All the required information about the variable names and the parameters of send/receive constructs are also recorded.

3. **Generating Edges.**

This phase connects the basic blocks generated in the previous phase with the appropriate edges. The edges are classified into three categories, sequential, parallel, and synchronization edges. Sequential edges indicate a possible flow from a block to another one. This type of edges is used to connect the basic blocks within each process as the ordinary sequential flow edges. Parallel edges indicate the parallel control flow as at process declaration and termination points. Synchronization edges represent the inter-process communication via send/receive operations.
Synchronization edges are generated by matching the parameters of call MPI_Recv( ) and call MPI_Send( ) recorded in the second phase. The output of this phase is the MPI-CFG shown in figure 2.

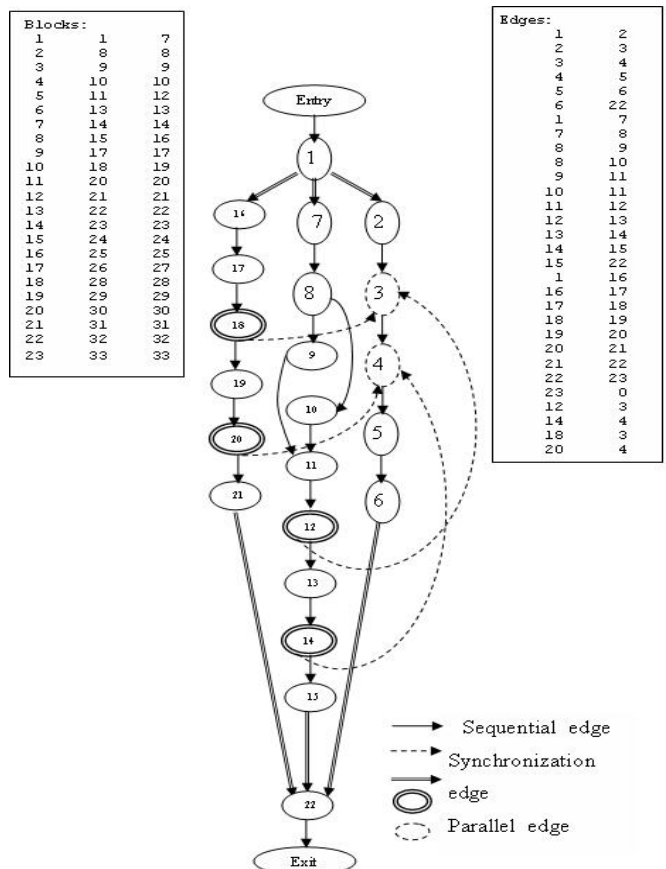

Figure 2. MPI Control Flow Graph for the MPI program

## 5. MPI Static Data Flow Analysis

Static data flow analysis is a technique for gathering information about the possible set of values calculated at various points in a sequential program. CFG is used to determine those parts of a program to which a particular value assigned to a variable might propagate. This can be done by generating two sets, $dcu(i)$ and $dpu(i, j)$ [13] for program variables.

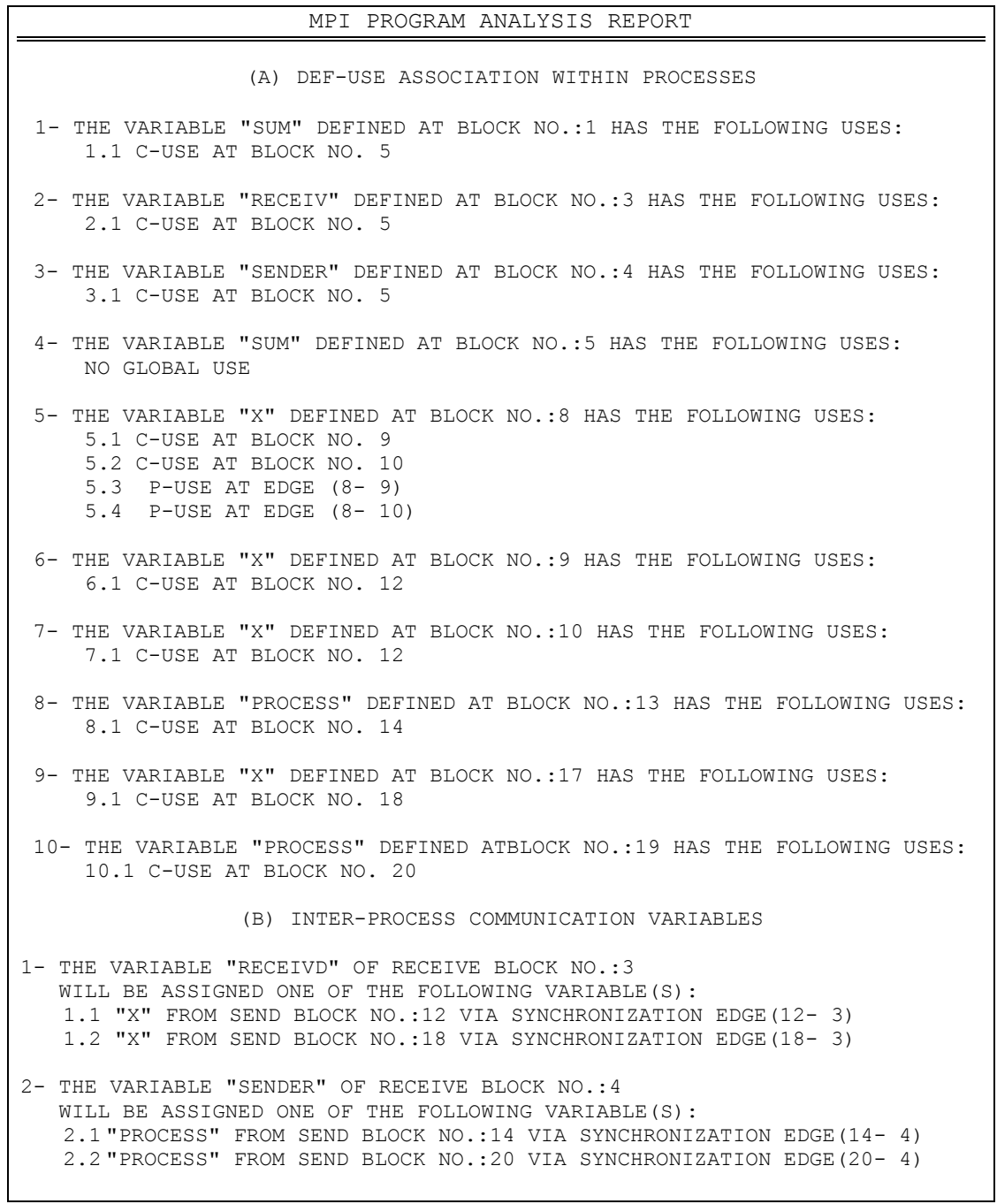

```
                    MPI PROGRAM ANALYSIS REPORT

               (A) DEF-USE ASSOCIATION WITHIN PROCESSES

 1- THE VARIABLE "SUM" DEFINED AT BLOCK NO.:1 HAS THE FOLLOWING USES:
      1.1 C-USE AT BLOCK NO. 5

 2- THE VARIABLE "RECEIV" DEFINED AT BLOCK NO.:3 HAS THE FOLLOWING USES:
      2.1 C-USE AT BLOCK NO. 5

 3- THE VARIABLE "SENDER" DEFINED AT BLOCK NO.:4 HAS THE FOLLOWING USES:
      3.1 C-USE AT BLOCK NO. 5

 4- THE VARIABLE "SUM" DEFINED AT BLOCK NO.:5 HAS THE FOLLOWING USES:
      NO GLOBAL USE

 5- THE VARIABLE "X" DEFINED AT BLOCK NO.:8 HAS THE FOLLOWING USES:
      5.1 C-USE AT BLOCK NO. 9
      5.2 C-USE AT BLOCK NO. 10
      5.3  P-USE AT EDGE (8- 9)
      5.4  P-USE AT EDGE (8- 10)

 6- THE VARIABLE "X" DEFINED AT BLOCK NO.:9 HAS THE FOLLOWING USES:
      6.1 C-USE AT BLOCK NO. 12

 7- THE VARIABLE "X" DEFINED AT BLOCK NO.:10 HAS THE FOLLOWING USES:
      7.1 C-USE AT BLOCK NO. 12

 8- THE VARIABLE "PROCESS" DEFINED AT BLOCK NO.:13 HAS THE FOLLOWING USES:
      8.1 C-USE AT BLOCK NO. 14

 9- THE VARIABLE "X" DEFINED AT BLOCK NO.:17 HAS THE FOLLOWING USES:
      9.1 C-USE AT BLOCK NO. 18

 10- THE VARIABLE "PROCESS" DEFINED ATBLOCK NO.:19 HAS THE FOLLOWING USES:
      10.1 C-USE AT BLOCK NO. 20

               (B) INTER-PROCESS COMMUNICATION VARIABLES

 1- THE VARIABLE "RECEIVD" OF RECEIVE BLOCK NO.:3
    WILL BE ASSIGNED ONE OF THE FOLLOWING VARIABLE(S):
     1.1 "X" FROM SEND BLOCK NO.:12 VIA SYNCHRONIZATION EDGE(12- 3)
     1.2 "X" FROM SEND BLOCK NO.:18 VIA SYNCHRONIZATION EDGE(18- 3)

 2- THE VARIABLE "SENDER" OF RECEIVE BLOCK NO.:4
    WILL BE ASSIGNED ONE OF THE FOLLOWING VARIABLE(S):
     2.1"PROCESS" FROM SEND BLOCK NO.:14 VIA SYNCHRONIZATION EDGE(14- 4)
     2.2"PROCESS" FROM SEND BLOCK NO.:20 VIA SYNCHRONIZATION EDGE(20- 4)
```

Figure 3. Data Flow Analysis Report

These two sets are necessary to determine the definitions of every variable in the program and the uses that might be affected by these definitions. The set $dcu(i)$ is the set of all variable definitions for which there are def-clear paths to their c-uses at node $i$ .

The $dpu(i, j)$ is the set of all variable definitions for which there are def-clear paths to their p-uses at edge $(i, j)$ [12]. Using information concerning the location of variable definitions and references, together with the "basic static reach algorithm" [5], the two sets can be determined. The basic static reach algorithm is used to determine the sets reach(i) and avail(i). The set reach(i) is the set of all variable definitions that reach node i . The set avail(i) is the set of all available variables at node i. This set is the union of the set of global definitions at node i together with the set of all definitions that reach this node and are preserved through it. Using these two sets, the sets $dcu(i)$ and $dpu(i, j)$ are constructed from the formula :

$$dcu(i) = reach(i) \cap c-use(i)$$
$$dpu(i, j) = avail(i) \cap p-use(i, j)$$

We applied this technique on the constructed MPI-CFG with some modifications to handle the nature of MPI programs. The output of the implemented technique is shown in figure 3.

## 6. Conclusion and Future Work

Unlike sequential programs, data flow analysis of MPI programs requires extra effort. Applying existing concepts, techniques and tools used to analyze sequential programs on MPI programs fails to report a correct program analysis. These techniques require some modifications to handle the SPMD nature of MPI programs. We have implemented a technique to extend the program CFG to represent the MPI programs MPI-CFG. The static analyzer uses the constructed graph to generate the program analysis report. We have implemented the techniques of building message basic blocks, constructing parallel edges, and also constructing synchronization edges represent send/receive constructs to produce a correct MPI program representation. In future, we hope to implement the construction of synchronization edges for all MPI inter-process communication constructs.